\documentstyle[12pt]{article}
\begin{document}

\vspace{0.5in}
\begin{center}
{\Large\bf New Universality Classes for Quantum Critical Behavior
}\\
\medskip
\vskip0.5in

\normalsize{\bf H.Hamidian$^{\rm a}$, G.Semenoff$^{~\rm b}$,
P.Suranyi$^{\rm a}$ , L.C.R.Wijewardhana$^{\rm a,c}$}
\smallskip
\medskip

{ \sl
$^a$Department of Physics,University of Cincinnati\\
Cincinnati,Ohio,45221, U.S.A.\\~~\\
$^b$Department of Physics, University of British Columbia\\
Vancouver,British Columbia,Canada,V6T 1Z1.\\~~\\
$^c$Institute of Theoretical Physics,Uppsala University,\\
S-751 08, Uppsala,Sweden \\
}\smallskip
\end{center}
\vskip 1.0in

\begin{abstract}
\vskip 0.25truein
We use the epsilon expansion to explore a new universality class of second
order quantum phase transitions associated with a four-dimensional Yukawa field
theory coupled to a traceless Hermitean matrix scalar field.  We argue that
this class includes
four-fermi models in $2<D<4$ dimensions with
$SU(N_F)\times U(N)$ symmetry and a $U(N)$ scalar, $SU(N_F)$
iso-vector 4-fermi coupling.
The epsilon expansion indicates that there is a second order phase transition
for  $N\geq N^*(N_F)$, where $N^*(N_F)\simeq.27N_F$ if
$N_F\rightarrow\infty$.\end{abstract}
\vfill\eject
\baselineskip=20pt

There are several scenarios in field theory where quantum critical
phenomena play a role.  They are intimately associated with
renormalization theory and are important in modifications of the
standard model
using technicolor \cite{atew} or models with composite Higgs fields
\cite{n}-\cite{cgs}.
Quantum critical phenomena are associated with phase transitions which
are achieved, typically at zero temperature, by tuning mechanical
parameters such as particle masses or coupling constants to critical
values.  Classical critical phenomena, on the other hand, occur at
non-zero temperature and are associated with phase transitions where
temperature or some other thermodynamic variable is the quantity which
must be tuned.

An important concept in the study of second order phase transitions is
universality. A phase transition is characterized by the number of
degrees of freedom which become massless at the critical point, as
well as the symmetry of the system and the symmetry breaking pattern
associated with the transition.  In classical critical phenomena, at
finite temperature, fermions decouple from infrared behavior because
of their half-odd-integer Matsubara frequency.  Thus, all universality
classes are represented by Bose field theories.  In quantum critical
behavior, fermions can be massless at a phase transition and therefore
they must be included in the effective low energy field theories which
represent universality classes.

A useful tool for the study of phase transitions is the Wilson-Fisher
epsilon expansion where the beta function is computed in a
$4-\epsilon$ dimensional effective field theory and second order phase
transitions are characterized by the infrared stable fixed points of
the renormalization group flow.  If the beta function exhibits such
fixed points and if the coupling constants have initial conditions
within their domain of attraction, there is a second order transition.
If no infrared stable fixed points exist then if there is a phase
transition it must be of first order, driven by
fluctuations~\cite{a}-\cite{y}.

The epsilon expansion has been used to show that the $O(N)$ vector
model in $4-\epsilon $ dimensions has a second order phase transition
with computable critical exponents \cite{amit}.  It represents a
universality class which contains the $O(N)$ nonlinear sigma model in
dimensions between 2 and 4.  The sigma model is renormalizable only in
the 1/N expansion \cite{arafeeva} and represents the continuum limit
of generalized magnetic systems.  It has also been argued
\cite{zj}-\cite{rwp1} that the four dimensional Yukawa-$\phi^4$ theory
in $4-\epsilon$ dimensions represents the universality class of the
Gross-Neveu model in dimensions between 2 and 4.  The latter is also
renormalizable only in the 1/N expansion ~\cite{rwp,s} and the
critical exponents have been computed to high order
\cite{kog}-\cite{vas}.  The $O(N)$ model has one relevant coupling
constant in the 4-dimensional theory, so the fixed point of order
$\epsilon$ is automatically infrared stable and the phase transition
is of second order.  When there is more than one relevant interaction
in the $4-\epsilon$ dimensional theory, and therefore a
multi-dimensional coupling constant flow, the generic situation is
that fixed points of the flow are not infrared stable and the phase
transitions are first order.  The classic example of this is the
massless scalar electrodynamics studied by Coleman and Weinberg
\cite{cw}.  There, in the two-dimensional plane of electromagnetic and
Higgs self-coupling there are no infrared stable fixed points.
Another example occurs in the complex matrix scalar field theory with
$SU(N_L)\times SU(N_R)$ symmetry which represents the universality
class of the finite temperature chiral transition in QCD \cite{pw}.
There are two coupling constants for the renormalizable interactions
${\rm tr}(M^{\dagger}M)^2$ and $({\rm tr}M^{\dagger}M)^2$ and when the
matrices are larger than $2\times 2$ the phase transition is first
order.  When the matrices are $2\times2$ the model is equivalent to a
vector theory with one coupling constant and the phase transition is
second order.  Similar reasoning has been used to argue that a
$4-\epsilon$-dimensional traceless Hermitean matrix scalar field
theory with $SU(N_F)$ symmetry, which also has two relevant couplings
${\rm tr}\phi^4$ and $({\rm tr}\phi^2)^2$, has a second order phase
transition only when $N_F=2$ and has a fluctuation induced first order
phase transition when $N_F>2$ ~\cite{p}.

In this Letter, we shall show that when a Yukawa coupling to fermions
is introduced to the Hermitean matrix theory, the existence of a
nontrivial Yukawa fixed point tends to make the fixed points of the
matrix theory infrared stable.  For a given $N_F$, we always find a
second order phase transition if the number of fermions in the
fundamental representation is large enough, asymptotically $N>.27N_F$
for large $N,N_F$, and first order otherwise.  We shall also argue
that the Yukawa-matrix model represents the universality class of a
second order quantum phase transition in a matrix-coupled four-fermi
theory in $2<D<4$ dimensions.  We use this equivalence to argue that,
for large enough $N$, the epsilon expansion is valid for epsilon of
order 1.

The Euclidean action for the $4-\epsilon$-dimensional Yukawa model is
\begin{eqnarray}
S&=&\int d^{4-\epsilon}x\Bigg( \bar\psi^a_\alpha\left(\delta_{\alpha\beta}
\vec\gamma\cdot\vec\nabla+\frac{\pi\mu^{\epsilon/2}y}{\sqrt{N_FN}}
\phi_{\alpha\beta}\right)
\psi^a_\beta\nonumber\\&+&
\frac{1}{2}{\rm tr}_F\vec\nabla\phi\cdot\vec\nabla\phi
+\frac{8\pi^2\mu^\epsilon}{4!}\left(\frac{g_1}{N_F^2}({\rm
tr}_F\phi^2)^2+\frac{g_2}{N_F}{\rm tr}_F\phi^4\right)\Bigg),
\label{gl}
\end{eqnarray}
where $a,b=1,\ldots N$,  $\alpha,\beta=1,\ldots,N_F$,  $\phi_{\alpha\beta}$ is
a traceless Hermitean $N_F\times N_F$ scalar matrix field.  $\psi_\alpha^a$ is
a complex Dirac spinor.
Here we have included all vertices which are compatible with the unitary
symmetry, $\phi\rightarrow U\phi U^{\dagger}$, $\psi\rightarrow U\psi$,
$\bar\psi \rightarrow \bar\psi U^{\dagger}$ with $U\in SU(N_F)$ and which are
renormalizable in 4 dimensions.  We have scaled the coupling constants so that
in the large $N_F$ limit planar graphs dominate and fermion loops are
suppressed and in the large $N$ limit the bubble graphs dominate.

The beta functions of this model for the case $y=0$ was computed to one-loop
order in \cite{p}.  The beta function for the model with $y\neq 0$ is readily
obtained to the same order:
\begin{equation}
\beta_1=-\epsilon g_1+\frac{N_F^2+7}{6N_F^2}g_1^2+\frac{2N_F^2-3}{3N_F^2}g_1g_2
+\frac{N_F^2+3}{2N_F^2}g_2^2+\frac{1}{2N_F}y^2g_1,
\end{equation}
\begin{equation}
\beta_2=-\epsilon
g_2+\frac{2}{N^2_F}g_1g_2+\frac{N_F^2-9}{3N_F^2}g_2^2-\frac{3}{8NN_F}y^4
+\frac{1}{2N_F}y^2 g_2,
\end{equation}
\begin{equation}
\beta_y=-\frac{\epsilon}{2} y+\frac{2NN_F+N_F^2-3}{16N_F^2N}y^3.
\end{equation}
Also at one-loop order, the scaling dimension of the scalar field is
\begin{equation}
\Delta_B= \frac{2-\epsilon}{2}+\frac{1}{8N_F}y^2+\ldots,
\end{equation}
and that of the fermion field is
\begin{equation}
\Delta_F=  \frac{3-\epsilon}{2}+\frac{1}{32}\frac{N_F^2-1}{N_F^2N} y^2+\ldots
\end{equation}

There will be a second order phase transition when the beta function has
infrared stable fixed points.  This occurs where $\beta_i(g)=0$ and where all
eigenvalues of the matrix $\partial \beta_i/\partial g_j$ are positive.  The
trivial fixed point $y_0^*=0$ of the Yukawa coupling always coincides with a
negative eigenvalue of the stability matrix.  It could therefore be ultraviolet
stable, but not infrared stable.  Thus, to find infrared fixed points, we take
the nonzero fixed point of the Yukawa coupling,
\begin{equation}
y^2_*=\frac{8N_F^2N}{2NN_F+N_F^2-3}\epsilon.
\label{yuk}
\end{equation}
This coupling is small when $\epsilon$ is small.

In the large $N$ limit, the beta functions have an infrared stable
fixed
point at Yukawa coupling (\ref{yuk}) and scalar self-couplings
\begin{equation}
g_1^*=-\frac{18(N_F^2+3)}{N^2}\epsilon+\ldots
{}~,~~
g_2^*=\frac{6N_F}{N}\epsilon+\ldots
\label{scal}
\end{equation}
In this regime, the scalar self couplings are small and the
one-loop perturbation theory is accurate.  The scaling dimensions of the
spinor and scalar fields can be computed to leading order in $\epsilon$  and
$1/N$ at the fixed points
\begin{equation}
\Delta_F= \frac{3-\epsilon}{2}+\frac{N_F^2-1}{8NN_F} \epsilon+\ldots,
\label{dyf}
\end{equation}
\begin{equation}
\Delta_B= 1+\ldots
\label{dyb}
\end{equation}
These critical exponents depend
on both $N$ and $N_F$ and therefore each value of these parameters specifies a
different universality class.

We have solved for the fixed points numerically for general values of $N$ and
$N_F$. The asymptotic form of the fixed points in  (\ref{yuk}) and (\ref{scal})
continued to lower  $N$ are depicted for a few values of $N_F$  in Fig. 1.
A critical value of $N$, $N^*(N_F)$, can be defined for every $N_F$, such that
an infrared stable fixed point exists at $N\geq N^*(N_F)$.
If $N_F\rightarrow\infty$ then $N^*(N_F)\simeq0.27N_F$. This asymptotic limit
is reached through values of $N^*(2)=34$, $N^*(4)=8$, $N^*(8)=4$, $N^*(80)=22$,
$N^*(800)=216$.

An example of a dynamical system whose universality class is described by
(\ref{gl}) is the four-fermi theory~\cite{ssw}
\begin{equation}
S=\int d^Dx\left(
\bar\psi_\alpha^a\vec\gamma\cdot\vec\nabla
\psi_\alpha^a-\frac{\lambda}{N}\bar\psi^a_\alpha T^A_{\alpha\beta}
\psi^a_\beta\bar\psi^b_\gamma T^A_{\gamma\delta}
\psi^b_\delta\right),
\label{model}
\end{equation}
where $2<D<4$, $\alpha,\beta,\dots=1,\ldots,N_F$,
$a,b,\dots=1,\ldots,N$ and  $\psi$ is a 4-component Dirac
spinor.
$T^A$ are generators of $SU(N_F)$ in the fundamental
representation, normalized so that $\rm tr T^AT^B=\delta^{AB}/2$.
This model has the same symmetries as (\ref{gl}): $C$, $P$,
$T$, as well as discrete chiral symmetry and global $SU(N_F)\times U(N)$.
 \cite{footnote}
The parity and discrete chiral transformation of the four component fermions
can be defined in such a way that the mass operator $\bar\psi\psi$ transforms
as a scalar  under parity and changes sign under a discrete chiral
transformation~\cite{jt}.  The phase transition that we shall study generates
fermion mass through
spontaneous breaking of the $SU(N_F)$ and discrete chiral symmetry.  The order
parameter is
$\hat\phi_{\alpha\beta}\equiv
(-\lambda/N)<\bar\psi^a_\alpha \psi^a_\beta
-\frac{1}{N_F}\delta_{\alpha\beta}\bar\psi^a_\gamma\psi^a_\gamma>$.

Four-fermi models with isoscalar couplings are known to be renormalizable in
the large $N$ expansion \cite{rwp,s,w,sw} when the coupling is tuned to a
critical value where a second order phase transition takes place. Existence of
a phase transition has also been demonstrated in the conformal bootstrap
approach in isoscalar models~\cite{cms}. Their arguments can be extended to the
model (\ref{model}) in a straightforward way.
We introduce an $N_F\times N_F$ auxiliary field, $\phi$,
\begin{equation}
S=\int d^Dx\left(
\bar\psi_\alpha^a(\delta_{\alpha\beta}\vec\gamma\cdot\vec\nabla
+\phi_{\alpha\beta})\psi_\beta^a+\frac{N}{2\lambda}{\rm
tr}_F\phi^2\right),
\label{flavor}
\end{equation}
where ${\rm tr}_F\phi=0$.
Here, the fermions can be integrated to obtain the scalar field theory
\begin{equation}
S_{eff}=-N{\rm TR}\ln\left(\vec\gamma\cdot\vec\nabla+\phi\right) +\int d^Dx
\frac{N}{2\lambda}{\rm tr}_F\phi^2,
\label{flavoreff}
\end{equation}
which can be studied perturbatively in the $1/N$ expansion. (Here
${\rm TR}$ means
trace in function space and over indices.)  The effective scalar propagator for
the large $N$ expansion,
\begin{eqnarray}
\Delta_{\alpha\beta\gamma\delta}(p)&\equiv&
<\phi_{\alpha\beta}(p)\phi_{\gamma\delta}(-p)>=
\left(\delta_{\alpha\delta}\delta_{\beta\gamma}-
\frac{1}{N_F}\delta_{\alpha\beta}\delta_{\gamma\delta}\right)
\nonumber\\&\times&
\frac{1}{N}\left( \frac{1}{\lambda}- \frac{1}{\lambda_c}-\frac{
4\Gamma(1-D/2)[\Gamma(D/2)]^2p^{D-2} }{ (4\pi)^{D/2}\Gamma(D-1)}\right)^{-1}.
\label{prop}
\end{eqnarray}
$\lambda_c $ is the critical coupling which is of order $\Lambda^{2-D}$ where
$\Lambda$ is the large momentum cutoff.  The propagator is scale invariant when
$\lambda$ is tuned to $\lambda_c$.  This is the value of the coupling at which
the phase transition takes place.
When $\lambda $ has this critical value, the scaling
dimensions of the scalar and spinor fields
are defined by the behavior of the fermion and scalar propagators,
\begin{eqnarray}
S_{\alpha\beta}(p)\sim& \delta_{\alpha\beta}
\frac{1}{-i\vec\gamma\cdot\vec p}~ \vert p\vert^{2\Delta_F+1-D},
\\
\Delta_{\alpha\beta\gamma\delta}(p)\sim&
\left(\delta_{\alpha\delta}\delta_{\beta\gamma}-
\frac{1}{N_F}\delta_{\alpha\beta}\delta_{\gamma\delta}\right)
\vert p\vert^{2\Delta_B-D},
\end{eqnarray}
respectively.
{}From this and computation of the spinor
self-energy to leading order in 1/N and (\ref{prop}), we deduce
\begin{equation}
\Delta_F=\frac{D-1}{2}
-\frac{\Gamma(D-1)(N_F^2-1)}{8N_FN\Gamma(1-D/2)[\Gamma(D/2)]^2
\Gamma(D/2-1)}+\ldots~~~,
\label{df}
\end{equation}
\begin{equation}
\Delta_B=1+\ldots~~~.
\label{db}
\end{equation}
If we set $D=4-\epsilon$ in (\ref{df}) and expand to first order in $\epsilon$,
the scaling dimension of the fermion in the four-fermi model is identical with
that in the Yukawa-matrix model at the infrared stable fixed point in
(\ref{dyf}).  Furthermore, the scaling dimension of the scalar field to leading
order in (\ref{db}) agrees with that in (\ref{dyb}).  We conjecture that this
equivalence holds in higher orders of perturbation theory and that the two
models are in the same universality class.

Symmetry breaking in the model (\ref{model}) can also be studied in the 1/N
expansion by computing the effective potential for $\hat\phi\equiv
\langle
\phi\rangle$ which to leading order in $N$ is readily obtained from
(\ref{flavoreff}) as
\begin{equation}
\lim_{N\rightarrow\infty}\frac{1}{N} V_{\rm eff}(\hat\phi)=
\left(\frac{1}{\lambda}-\frac{1}{\lambda_C}\right)\frac{1}{2}
{\rm tr}_F\hat\phi^2+2
\frac{\Gamma(-D/2)}{(4\pi)^{D/2}}{\rm tr}_F(\hat\phi^2)^{D/2}~~.
\label{tree}
\end{equation}
This potential clearly
exhibits a second order phase transition at $\lambda \rightarrow
\lambda_C$.  When $\lambda>\lambda_C$, the eigenvalues of $\hat\phi$
obtain an
expectation value with equal magnitude and undetermined sign.  This
symmetry breaking is consistent with the fact that ${\rm
tr}\hat\phi=0$ only if $N_F$ is even and the symmetry breaking
pattern
is $SU(N_F)
\rightarrow SU(N_F/2)\otimes SU(N_F/2)$.
(If $N_F$ is odd,
we can show for small values of $N_F$ that the symmetry breaking
pattern with $(N_F-1)/2$ positive, $(N_F-1)/2$ negative and one zero
eigenvalue is preferred. In the following we shall avoid this complication by
assuming that $N_F$ is even.)
If we assume this
symmetry breaking pattern, it is straightforward to compute the
leading correction to (\ref{tree}) when $D=3$ and
$\lambda=\lambda_C$.
The 1-loop correction to the effective action is given by
\cite{ssw}
\begin{equation}
V_{\rm eff}(\hat\phi)=
\frac{1}{3\pi}N_FN\vert\hat\phi\vert^{3}\left(
1-\frac{4}{\pi^2}\frac{N_F-2}{N}\ln\vert
\hat\phi\vert/\mu+\ldots\right)
\approx \frac{1}{3\pi}NN_F\vert\hat\phi\vert^{3-
\frac{2}{\pi^2}\frac{N_F-2}{N} }.
\label{loop}
\end{equation}
The logarithmic corrections which are typical of next to leading order
in the effective potential (and which can in some cases lead to a
Coleman-Weinberg instability ) exponentiate to give the scaling
dimension of the potential~\cite{ssw}.  Thus we have established that
if $N$ is large enough, the model has a second order phase transition.
For large $N_F$ the critical number of colors is $N^*=.27N_F$. At
$N_F=2$ $N^*=34$, at $N_F=4$ $N^*=8$, at $N_F=200$ $N^*=55$. The
behavior of the critical number of colors as a function of the number
of flavors is consistent with results obtained in equivalent four
fermi interaction theories~\cite{ssw}.  It is interesting to speculate
that the upper critical value of $N_F\sim3.7N$ for existence of a
chiral transition in the four-fermi theory is related to the upper
critical $N_F\sim{\rm~ const.}\cdot N_{\rm color}$ for existence of
chiral symmetry breaking in 2+1 dimensional QCD~\cite{ap2}. A
correspondence of this kind was the intention of ref.\cite{p}.
However, if we naively identify U(N) with color, that theory has
either different symmetry or different degrees of freedom: If the QCD
is confining, U(N) is absent and if it is not confining, massless
gluons should contribute to the critical behavior.  Furthermore, in
gauge theories at zero temperature, once $N_F$ and $N$ are given,
either the vacuum is symmetric or the symmetry is spontaneously
broken.  There is no mechanical parameter that drives the symmetry
breakdown unlike at finite temperature.  Suppose an additional
interaction such as the four-fermi coupling that we have considered in
this paper is included in the theory.  Then this type of effective low
energy field theory might be applicable for the gauged Nambu-Jona
Lasinio model and our analysis would apply close to the
Nambu-Jona-Lasinio fixed point where the gauge interaction is weak
{}~\cite{atw1}.  But in the region of phase space where the gauge
interaction is strong a more refined analysis is needed.  It has been
seen that in this region the effective potential exhibits non-analytic
behavior~ \cite{atw2}.  This issue warrants further study.

\section*{ACKNOWLEDGEMENTS}

 The work of G. S. is supported by the Natural Science and
Engineering
Research Council of Canada.  The work of H. H., P. S. and of L. C. R. W. is
supported in part by the United States Department of Energy under
grant no.  DE-FG02-84ER40153. The work of L.C.R.W. is supported by
Uppsala University. He thanks A.Niemi for hospitality at
the Theoretical Physics Department at Uppsala.

\noindent
{\bf Figure Captions}

\noindent
Figure 1: Fixed point values of the coupling constants $g_1$ and $g_2$ as $N$
is varied from infinity where $(g_1^*,g_2^*)=(0,0)$ to its lower critical value
which is $N^*=8$ for $N_F=4$ (bottom curve), $N^*=4$ for $N_F=8$ (middle
curve) and $N^*=22 $ for $N_F=80$ (top curve).


\begin{thebibliography}{10}
\small
\addtolength{\itemsep}{-6pt}
\bibitem{atew}T. Appelquist, T. Takeuchi, M. Einhorn and L.C. R. Wijewardhana,
Phys. Lett. {\bf B220}, 223 (1989).
\bibitem{n}Y. Nambu, in New Trends in Physics, Proceedings of the XI
International Symposium on Elementary Particle Physics, Kazimierz, Poland,
1988, edited by Z. Adjuk, S. Pokorski and A. Trautmann (World Scientific,
Singapore, 1989).
\bibitem{my}V. Miransky, M. Tanabashi  and K. Yamawaki, Phys. Lett. {\bf 221},
1043
(1989); Mod. Phys. Lett. {\bf A4}, 1043 (1989).
\bibitem{bhl}W. Bardeen, C. Hill and M. Lindner, Phys. Rev. D 41, 1647 (1990).
\bibitem{bh}W. Bardeen, C. Hill and Jungnickel, Phys. Rev. D49, 1437 (1994).
\bibitem{cgs}R. Chivukula, A. Cohen and K. Lane, Nucl. Phys. B343, 5543 (1990);
R. Chivukula, M. Golden and E. Simmons, Phys. Rev. Lett. 70, 1587 (1993).
\bibitem{a}I. Iacobson and D. Amit, Ann. Phys. 133, {\bf 57} (1981).
\bibitem{cw}S. Coleman and E. Weinberg,  Phys. Rev. {\bf  D7}, 1888
(1973).
\bibitem{y}H. Yamagishi, {\sl Phys. Rev.} {\bf D23}, 1880
(1981).
\bibitem{amit}For a review and references, see D. Amit, {\it Field Theory, the
Renormalization Group and Critical Phenomena}, McGraw-Hill, 1978.
\bibitem{arafeeva}I. Araf'eva, E. Nussimov and S. Pacheva, Comm. Math. Phys.
{\bf 71},
213 (1980).
\bibitem{zj}J. Zinn-Justin, Nucl. Phys. {\bf B367}, 105 (1991).
\bibitem{k}A. K\"arkk\"ainen, R. Lacaze, P. Lacock and B. Peterson, Nucl. Phys.
{\bf 415}, 781 (1994).
\bibitem{rwp1}B. Rosenstein, H.-L. Yu and A. Kovner, Phys. Lett. {\bf B314},
381
(1993).
\bibitem{rwp}B. Rosenstein, B. Warr and S. Park, Phys. Rev. Lett. {\bf 62},
1433
(1989).
\bibitem{s}C. de Calan, P.A. Faria de Verga, J. Magnen and R. Seneor. Phys.
Rev. Lett. {\bf 66}, 3233 (1991).
\bibitem{kog}S. Hands, A. Kocic and J. Kogut, Ann. Phys. {\bf 224}, 29 (1993).
\bibitem{gr}J. Gracey, Phys. Lett. B342, 297 (1995); Phys. Rev. {\bf D50}, 2840
(1994); Z. Phys. C61, 115 (1994); Int. Jour. Mod. Phys. {\bf A9}, 567 (1994)..
\bibitem{vas} A.N. Vasilev, S.E. Derkachev, N.A. Kivel, A.S. Stepanenko, Theor.
Math. Phys. {\bf 94}, 127-136 (1993). (Teor. Mat. Fiz. {\bf 94},
179-192 (1993)): N.A. Kivel, A.S. Stepanenko, A.N. Vasilev,
Nucl.Phys. {\bf B424}, 619-627 (1994).
\bibitem{pw}R. Pisarski and F. Wilczek, Phys. Rev. {\bf D29}, 1222 (1984);
F. Wilczek, Nucl. Phys.
{\bf A566}, 123c-132c (1994); Int.J.Mod.Phys. {\bf D63}, 80 (1994);
Int.J.Mod.Phys. {\bf A7}, 3911-3925 (1992).
\bibitem{p}R. Pisarski, Phys. Rev. {\bf D44}, 1866 (1991).
\bibitem{ssw}G. Semenoff, P. Suranyi and L. C. R. Wijewardhana, Phys. Rev.
{\bf D50},
1060 (1994).
\bibitem{footnote} In $D=3$ the basic fermion has two components. In the
present paper, symmetry considerations force us to use four-componenent
spinors. Note that. with four-component spinors the kinetic term in
(\ref{model})
has $U(2N_F)$ symmetry, which is broken to $U(N_F)\times Z_2$, where $Z_2 $
is the discrete chiral symmetry, by the interaction term. Some three
dimensional theories cannot be ananlyzed using the epsilon expansion.
\bibitem{jt}R. Jackiw and S. Templeton, Phys. Rev. {\bf D23}, 2291 (1981).
\bibitem{w}K. Wilson, {\sl Phys.Rev.} {\bf D14}, 2911 (1974).
\bibitem{sw}G. W. Semenoff and L. C. R. Wijewardhana, Phys. Rev. Lett. {\bf
63},
 2633, (1989);  Phys. Rev.
{\bf D45}, 1342 (1992).
\bibitem{cms}W. Chen, Yu. Makeenko and G. Semenoff, Ann. Phys. (N.Y.) {\bf
288},
341 (1991).
\bibitem{atw1}T. Appelquist, J. Terning and L. C. R. Wijewardhana, Phys. Rev.
{\bf D44}, 871 (1991).
\bibitem{atw2}T. Appelquist, J. Terning and L. C. R. Wijewardhana, hep-ph
940320, submitted to Phys. Rev. Lett. 1994.
\bibitem{ap2}T. Appelquist and D. Nash, Phys. Rev. Lett. {\bf 64}, 721 (1990).
\end{thebibliography}
\end{document}